\documentclass{emulateapj}
\usepackage{natbib}
\newcommand{\re}{$R_{\rm e}$}
\newcommand{\lsun}{L$_\odot$}
\newcommand{\msun}{M$_\odot$}

\newcommand{\etal}{~et al.~}
\newcommand{\kms}{km s$^{-1}$~}
\shorttitle{A massive dark halo in a UDG}
\shortauthors{Beasley et al.}

\begin{document}

\title{An overmassive Dark Halo around an Ultra-diffuse Galaxy in the Virgo Cluster}

\author{Michael A.\ Beasley\altaffilmark{1,2},
  Aaron J.\ Romanowsky\altaffilmark{3,4},
  Vincenzo Pota\altaffilmark{5},
  Ignacio Martin Navarro\altaffilmark{1,2,4},
  David Martinez Delgado\altaffilmark{6},
  Fabian Neyer\altaffilmark{7},
  Aaron L.\ Deich\altaffilmark{3}}
\email{beasley@iac.es}

\altaffiltext{1}{Instituto de Astrofisica de Canarias, Calle Via Lactea, La Laguna, Tenerife, Spain}
\altaffiltext{2}{University of La Laguna. Avda. Astrof\'isico Fco. S\'anchez, La Laguna, Tenerife, Spain}
\altaffiltext{3}{Department of Physics and Astronomy, San Jos\'e State University, San Jose, CA 95192, USA}
\altaffiltext{4}{University of California Observatories, 1156 High Street, Santa Cruz, CA 95064, USA}
\altaffiltext{5}{INAF - Osservatorio Astronomico di Capodimonte, Salita Moiariello, 16, I-80131 Napoli, Italy}
\altaffiltext{6}{Astronomisches Rechen-Institut, Zentrum f\"{u}r Astronomie der Univer-
  sitat Heidelberg, M\"{o}nchhofstr. 12-14, 69120 Heidelberg, Germany}
\altaffiltext{7}{ETH Zurich, Institute of Geodesy and Photogrammetry, 8093 Zurich, Switzerland}

%% Abstract
% ===========================================================================
\begin{abstract}
  Ultra diffuse galaxies (UDGs) have the sizes of giants
  but the luminosities of dwarfs. A key to understanding their origins comes from their
  total masses, but their low surface brightnesses ($\mu(V) \geq$ 25.0) 
  generally prohibit dynamical studies. 
  Here we report the first such measurements for a UDG (VCC~1287 in the Virgo cluster),
  based on its globular cluster system dynamics and size.
  From 7 GCs we measure a mean systemic velocity $v_{\rm sys}$ = 1071$^{+14}_{-15}$ km/s,
  thereby confirming a Virgo-cluster association. We measure a velocity dispersion
  of 33$^{+16}_{-10}$ km/s within 8.1 kpc, corresponding to an enclosed mass of 
  $(4.5 \pm 2.8)\times10^{9}$ \msun\ and a $g$-band mass-to-light ratio of  
  $(M/L)_g = 106^{+126}_{-54}$. From the cumulative mass curve, along with the GC numbers,
  we estimate a virial mass of $\sim8\times10^{10}$~\msun, yielding
  a dark-to-stellar mass fraction of $\sim3000$. We show that this UDG is an outlier
  in $M_{\rm star}$--$M_{\rm halo}$ relations, suggesting extreme stochasticity in relatively massive
  star-forming halos in clusters. Finally, we discuss how counting GCs offers an efficient route to
  determining virial masses for UDGs.
 \end{abstract}

%% Keywords
% ===========================================================================

%% Sections
% ===========================================================================
\section{Introduction}

Deep imaging surveys of the Fornax (Mu\~noz\etal2015), Virgo 
(Ferrarese\etal2012; Mihos\etal2015), Coma (van Dokkum\etal2015a)
and the Pisces-Perseus supercluster (M\'artinez-Delgado\etal2016)
are revealing substantial populations of faint systems that were hidden from shallower surveys.
Perhaps most startling have been the results in Coma.  van Dokkum\etal(2015a) identified 47
``ultra diffuse galaxies'' (UDGs) consisting of seemingly
quiescent stellar populations with  characteristic luminosities, sizes, and central
surface brightnesses (SB)
of $L_g\sim0.1-2.5\times10^8 L_\odot$, \re $\sim1.5-4.6$ kpc, and $\mu_g\sim25$ mag arcsec$^{-2}$.
That is, these are galaxies with sizes similar to that of the Milky Way ($\sim2.15$ kpc scalelength, or
$\sim3.6$ kpc \re; Bovy\etal2013), but stellar luminosities
more akin to dwarfs.
Koda\etal(2015) identified a further $\sim1000$ UDGs in Coma
from deep Subaru imaging. The inference is that these galaxies may be
only the tip of the iceberg of ultra-faint stellar systems in clusters.

There are several possible formation pathways for UDGs.
They may be descendents of ``normal'' galaxies that have been altered
within the cluster tidal field (Gnedin 2003).
Alternatively, they may be  ``tidal dwarfs'', systems that were formed
during galaxy interactions and then lost to the cluster potential to exist in a transient,
free-floating phase (Bournaud\etal2007).
A third possibility is that they are ancient, remnant systems, perhaps either a
species of ``peculiar dwarf'' or ``failed giant,'' depending upon their total
masses. This last category could explain the survival of UDGs in cluster environments,
but would also imply that they are among the most dark-matter (DM) dominated galaxies in the universe,
with mass-to-light ratios of $>50$ within only 2~\re (van Dokkum\etal2015b).

Determining total masses for UDGs is clearly
a priority for understanding their formation and evolutionary
pathways. Unfortunately, due to their low surface brightnesses, obtaining
galaxy integrated spectra of sufficient SNR for robust
stellar velocity dispersions is extremely challenging with current
instrumentation.

Taking a different approach, in this {\it Letter}  we have obtained the first mass measurements
of a UDG, via the dynamics and statistical properties of its globular cluster (GC) system. 

\section{Data}
\subsection{Galaxy and globular cluster photometry}

\begin{figure*}
\epsscale{1.0}
\plotone{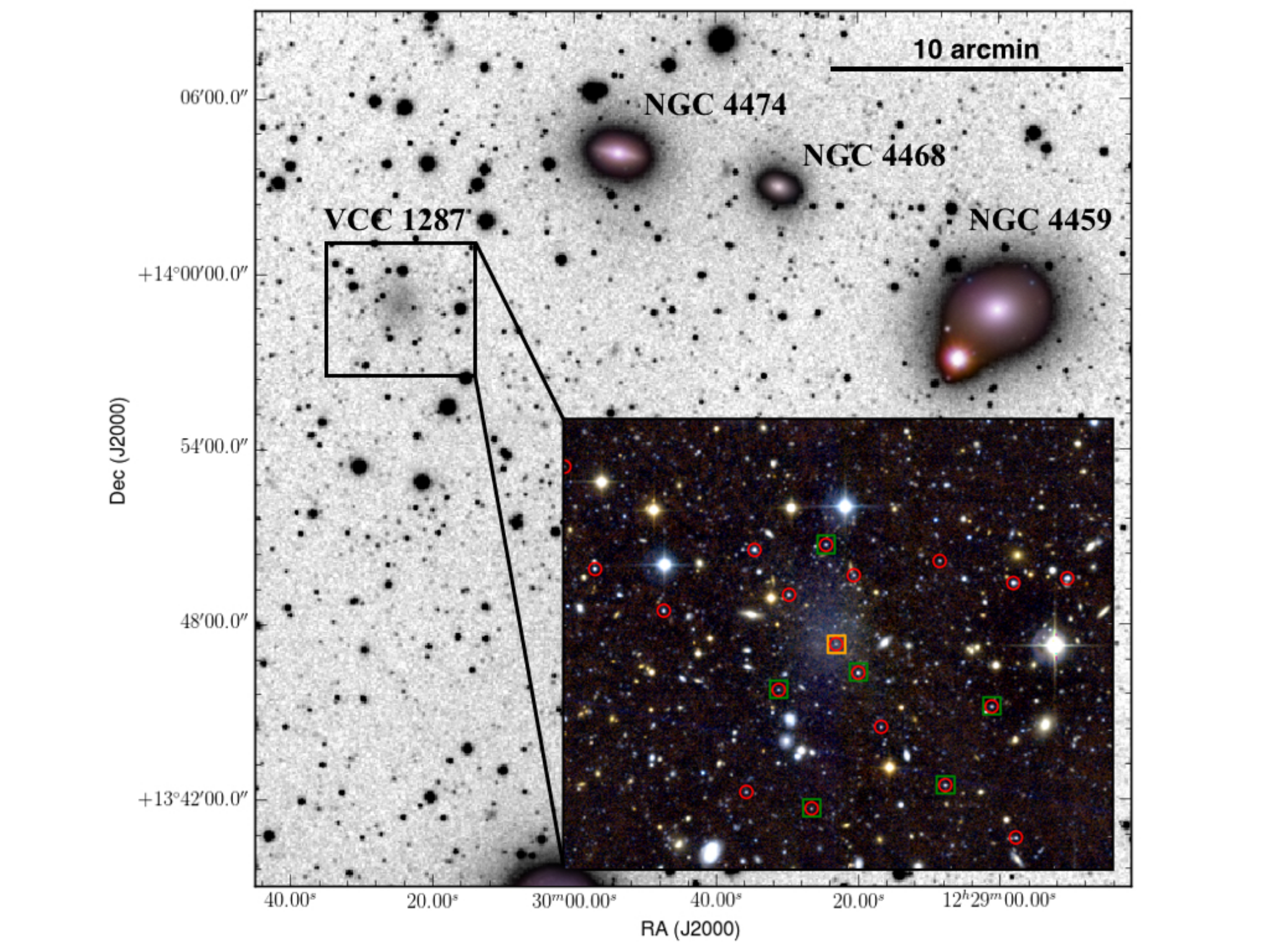}
\caption{The environment surrounding VCC~1287,
  observed with a 0.1-meter aperture at f/5.7 Borg ED101 apocromatic refractor from the Antares Observatory
  in north-eastern Switzerland.
  The subplot is a zoom-in $gri$ color-composite image of VCC~1287 from CFHT/MegaCam.
  %rebinned with a pixel scale of 0.374 arcsec px$^{-1}$. 
  GC candidates are circled in red.
  The green and orange boxes identify confirmed GCs and the confirmed nucleus of the
  galaxy, respectively. North is Up, East is Left.
}\label{fig:coverage}
\end{figure*}

\begin{figure*}
\epsscale{0.8}
\plotone{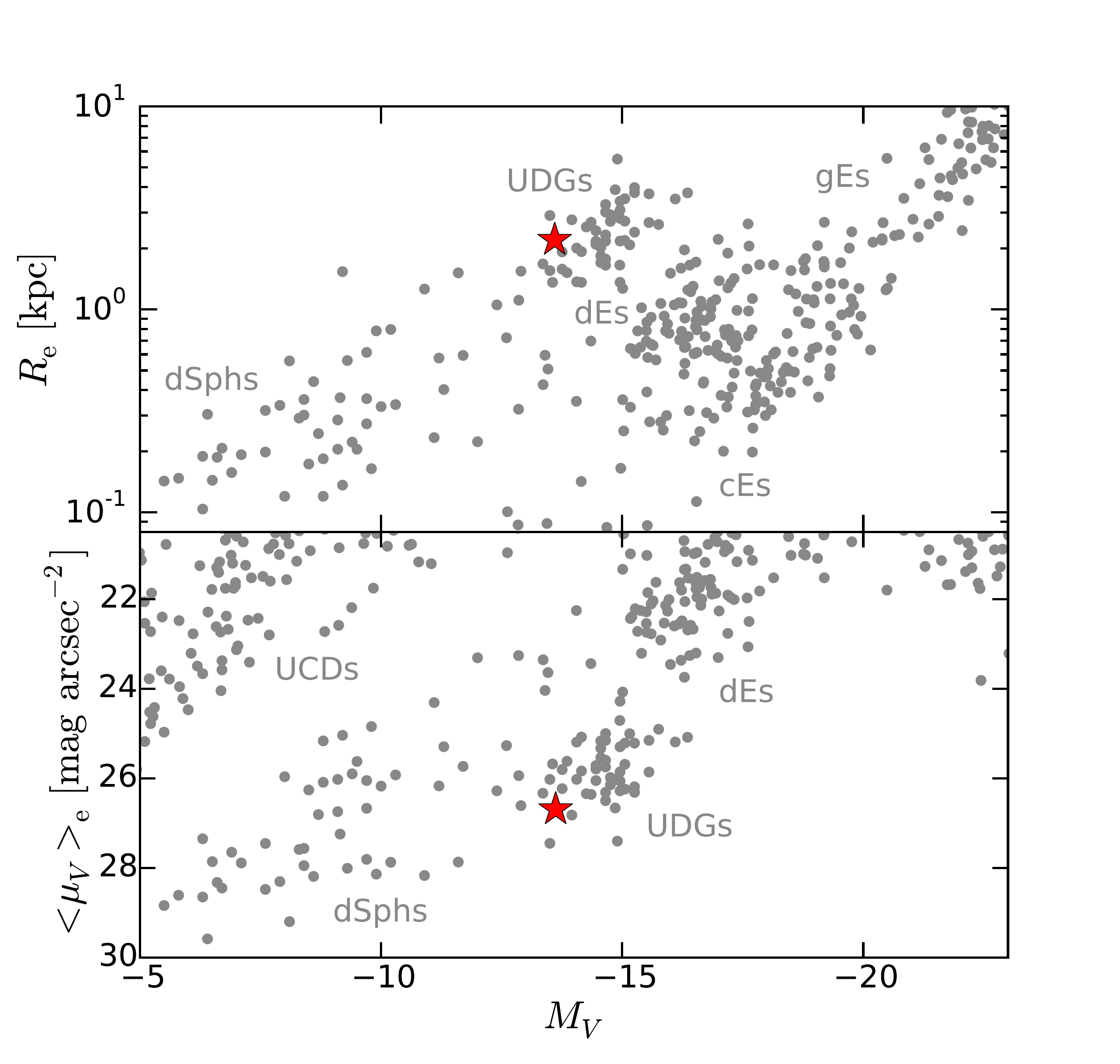}
\caption{Relations between size (circularized \re), mean surface brightness, and
  absolute magnitude for hot stellar systems (after van Dokkum\etal2015b).
  The gray points show distance-confirmed objects
  from the compilation of Brodie\etal(2011), with updates in
  ${\tt http://sages.ucolick.org/spectral\_database.html}$.
  Also included are UDGs from van Dokkum\etal(2015a) and Mihos\etal(2015).
  The red star marks VCC~1287. }\label{fig:size}
\end{figure*}

We identified UDG candidates in Virgo
using deep images taken with a 10-cm apocromatic refractor and archival
CFHT/MegaCam imaging. In particular, we selected one galaxy, VCC~1287, 
with all the characteristics of a UDG (Figures~\ref{fig:coverage} and \ref{fig:size}).
We analyzed its photometry from
%Photometry of VCC~1287 and its GC system was measured from the MegaCam images.
%MegaCamhas a pixel scale of 0.187 arcsec pix $^{-1}$.
1 deg$^2$ MegaCam images in five bands ($u$, $g$, $r$, $i$, $z$),
from the Canadian Astronomy Data Center. All images had been reduced and calibrated
using the Megapipe image stacking pipeline (Gwyn 2008).

The galaxy light was modeled using the software MegaMorph/GALFITM (Vilka\etal2008), which performs
simultaneous multi-band 2D fitting. All five MegaCam bands were first rebinned to
0.364 arcsec pix$^{-1}$ and then fitted using a single S\'ersic function.
Magnitudes and \re~were allowed to vary in different bands as a linear function of
wavelength, whereas the S\'ersic index $n$, the minor-to-major axis ratio $b/a$, and the position
angle $PA$ were constant in all images. From GALFITM, we obtained $n = 0.8$,  $b/a= 0.8$, and $PA = 36^\circ$.

We adopt a Virgo-cluster distance modulus of $m-M$ = 31.1 (Mei\etal2007).
We measure a mean, circularised \re across five photometric bands of
 \re$= 30.2\arcsec \pm 1.8\arcsec$  ($2.4 \pm 0.1$ kpc),
 and obtain a mean SB of $\mu(g,0)$ = 26.7 mag arcsec$^{-2}$,
 $(g-i)_0=0.83$ and $M_{g}=-13.3$.
By comparison, the UDGs identified by van Dokkum\etal(2015a) have
\re$=1.5-4.6$ kpc, $\mu(g,0)\sim25$ mag arcsec$^{-2}$, $(g-i)_0\sim0.8$
and $M_{g}\sim-14$ (Figure~\ref{fig:size}). VCC~1287 has already been identified as
a faint, extended system by Binggeli\etal(1985) in the Virgo Cluster Catalog (VCC),
although no redshift has been published.
Indeed, Binggeli\etal(1985) classified VCC~1287 as a dwarf irregular galaxy
``{\it .. of very large size and low surface brightness.}''

%The MegaCam imaging suggests a relatively rich GC system
%about VCC~1287, manifest as an excess of point sources with the
%correct colors and magnitudes for GCs at the distance
%of the Virgo cluster ({\bf S1.1}; Fig.~\ref{fig:coverage}).

GCs were identified exploiting all five MegaCam bands. A catalog of sources was extracted
using SExtractor (Bertin 1996). Only objects detected in all five bands were considered, selecting
point-sources by imposing the $i$-band {\textsc class star} parameter of SExtractor to
be $>0.5$. This removes most extended objects from the catalog, owing to
the excellent (0$\farcs$6) seeing of the images. We selected as GC candidates all point-sources
within the color ranges: $0.6 < (g-z) < 1.2, 1.1 < (u-g) < 1.8, 0.7 < (g-i) < 1.0$.
We selected a fairly narrow color range in order to minimize contamination
from background sources (in particular red background galaxies). However, these
colour ranges are consistent with the distributions seen in other studies of Virgo dwarf galaxies
(e.g., Beasley\etal2006; Peng\etal2006).
In order to exclude ultra-compact galaxies, we also imposed a magnitude cut, $i>19.5$ mag, equivalent
to the magnitude of the Milky Way GC $\omega$~Cen at the distance of Virgo.

To determine the total number of GCs in VCC~1287, we constructed the surface density profile of the
GC system and found it to extend to $\sim175$\arcsec($ \sim13$ kpc), beyond which is a
constant background with 0.2 objects per arcmin$^2$.
Next, all point-sources within 175\arcsec\ of the galaxy center and satisfying our selection criteria
were flagged as GCs -- returning 18 candidates.
Star and galaxy contamination was calculated by random placement of 1000
circles with 175\arcsec\ radii in the 1 deg$^2$ field surrounding VCC~1287.
We found a mean of 6 objects expected within such a circle,
with a standard deviation of 2.5 counts using our selection criteria, with
no sky apertures with $\geq18$ objects. The identification of 18 GC candidates
represents an overdensity with respect to the background level at
$\sim 5-\sigma$ significance.
Subtracting this contamination rate, left 11 GC
candidates down to the GC turn-over-magnitude ($i=23.1$, using $I$-band 
values of Kundu \& Whitmore (2001), and an $i$-band transformation from Faifer\etal(2011).

Assuming that the GC luminosity function is bell-shaped
Harris\etal(2000), we doubled their number to obtain  a total GC population
of $22 \pm 8$ GCs, where the uncertainties come from the quadrature
sum of the poisson uncertainties and the background contribution.

This may not sound remarkable, but when normalized to the
host galaxy magnitude using the ``specific frequency'' (S$_{\rm N}$; Harris \& van den Bergh 1981),
we find VCC~1287 has S$_{\rm N} = 80 \pm 29$. This is a significant overabundance
of GCs for VCC~1287's stellar luminosity (Figure~\ref{fig:cumMASS}).
%We return to this important point shortly.

\subsection{Globular cluster spectroscopy}

GC candidates were selected for multi-object
spectroscopy with the OSIRIS instrument on the Gran Telescopio de Canarias (GTC) in La Palma. 
%These candidates were re-identified in OSIRIS $r$-band pre-imaging for slitmask fabrication.
% using the OSIRIS mask-making software.
We observed this spectroscopic mask of VCC 1287 centered on
co-ordinates $\rm{RA}(J2000)=12^{\rm{h}}~30^{\rm{m}}~23.65^{\rm{s}}$,
$\rm{Dec.}~(J2000)=+13^{\rm{o}}~56^{\rm{m}}~46.3^{\rm{s}}$ during the nights
of 2015 June 19 and 20, under Director's Discretionary Time.

We placed slits (slitlength typically 10$\arcsec$, slitwidth 1$\farcs$2)
on 8 good candidates and the rest on lower priority objects.
The mask was observed for 4 hours in $\sim$0$\farcs$8
seeing and clear conditions. We used a 2500I grating centered on the NIR calcium-triplet
(CaT) region. To optimise sky subtraction, we nodded the objects along the
slit in an A--B pattern.
%thereby allowing us to subtract the sky in the same physical
%pixel as the object was observed (see, e.g., \cite{Carrera2007}).

\begin{figure*}
\epsscale{0.6}
\plotone{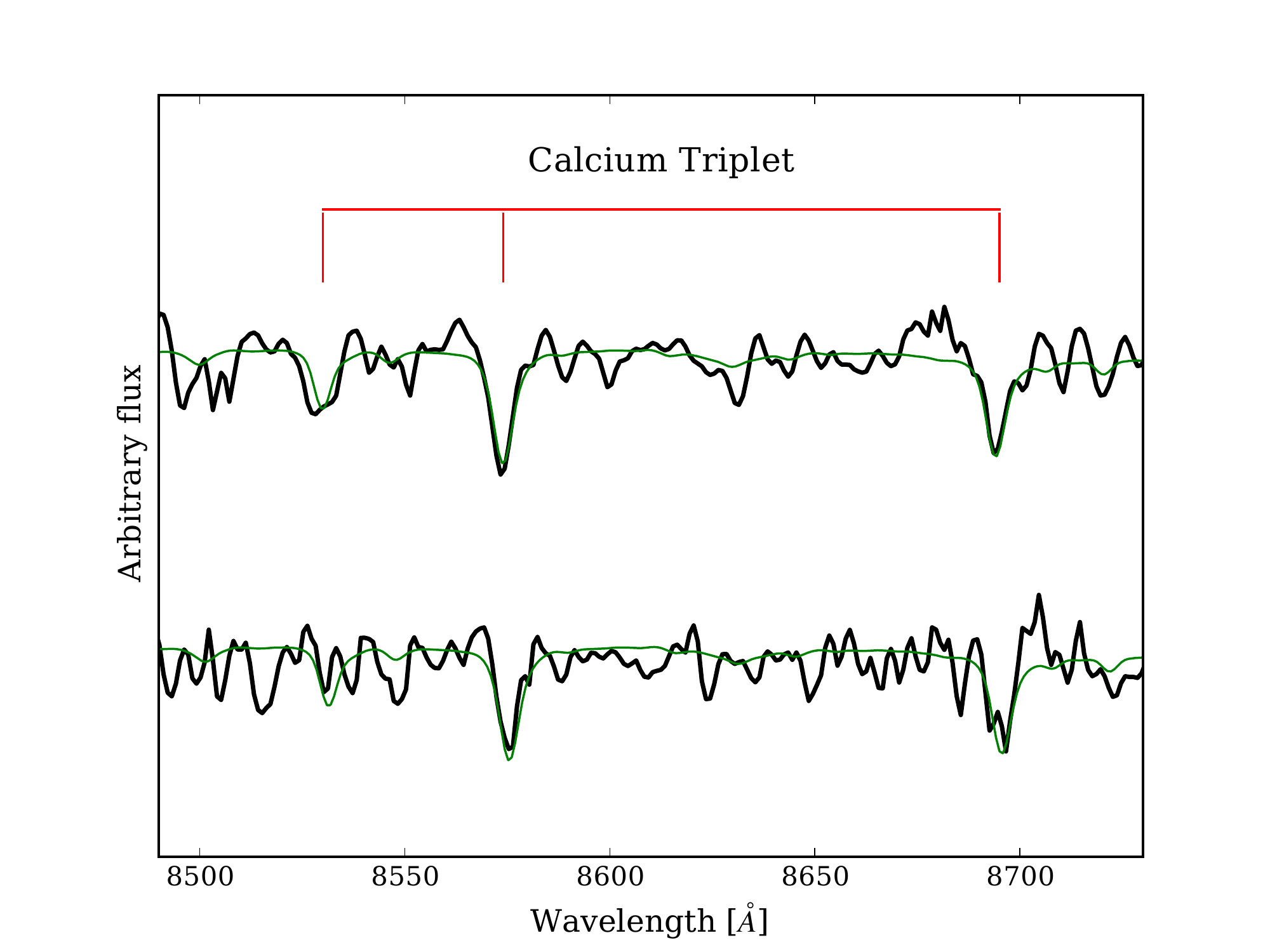}
\caption{Example CaT spectra of globular clusters around UDG VCC~1287.
  The top spectrum (GC16) has the highest SNR$\sim$25, the bottom spectrum (GC14)
  has one of the lowest (SNR$\sim$7). The positions of the CaT lines are indicated.
  The green lines show  the best-fitting spectral model templates, as determined from
  cross-correlation, from the Vazdekis\etal(2003) CaT models.
  The spectra have been divided by a low-order polynomial fit to the continuum.}\label{fig:CaT}
\end{figure*}

Data were reduced with IRAF. Individual 2-D spectra were cut from the
CCD frame, bias-subtracted, and divided by dome flat-fields. Corresponding HgCdArXe arcs
were cut out and a 2-D wavelength calibration and distortion map produced. Typical
residuals were $\sim$0.05~\AA. Difference spectra
were produced by subtracting the observed spectra in an A--B, B--A pattern.
These spectra were then wavelength calibrated, rectified in 2-D, extracted,
and combined as 1-D spectra for analysis.
%The spectra were not flux-calibrated.
The SNR of the spectra ranges from 7 to 25~\AA$^{-1}$ The spectral
resolution is $\sim5.5$~\AA\ (FWHM).
%measured from the arc line spectra.
Examples of the GC CaT spectra are shown in Fig~\ref{fig:CaT}.

Velocities were measured via fourier cross-correlation against a wide-range of model
templates (Vazdekis\etal2003), using FXCOR in IRAF.
For robust velocities we required a relative cross-correlation peak-height $>3$,
and that at least two of the three CaT lines were visible in our spectra (Strader\etal2011).
We estimated velocity uncertainties using  Monte Carlo simulations.
We degraded high-SNR CaT models to the resolution and the range of SNR of our spectra and also
mimicked sky-subtraction residuals by adding cosmic rays to the spectra with FWHM corresponding to that
of our spectral resolution. We generated 50 such spectra per SNR bin with a random seed
and measured velocities with FXCOR. We took the standard deviation on the mean of these velocities as
the typical velocity uncertainty for that SNR. For our spectroscopic sample,
we obtain velocity uncertainties are $\sim$20~\kms.

\begin{deluxetable*}{lccccccclc}
\tablecolumns{10}
%\tablewidth{0pc}
\tablecaption{Data for VCC~1287 GCs}
\tablehead{
\colhead{ID} & \colhead{RA(J2000)} & \colhead{DEC(J2000)}  &\colhead{$u$} & \colhead{$g$} &  \colhead{$r$} & \colhead{$i$} &\colhead{$z$} & \colhead{R} & \colhead{RV} \\
\colhead{} & \colhead{(deg)} & \colhead{(deg)} & \colhead{(AB mag)} & \colhead{(AB mag)} & \colhead{(AB mag)} & \colhead{(AB mag)} & \colhead{(AB mag)} & \colhead{($''$)} & \colhead{(\kms)} \\
}
\startdata 
GC10 & 187.6058982 & 13.9545477 & 25.34$\pm$0.08 & 23.73$\pm$0.04 & 23.28$\pm$0.04 & 22.80$\pm$0.04 &  22.57$\pm$0.05 & 93.9  & 1071$\pm$17\\
GC11 & 187.5842242 & 13.9581670 & 23.97$\pm$0.04 & 22.78$\pm$0.02 & 22.23$\pm$0.02 & 22.00$\pm$0.02 &  21.75$\pm$0.03 & 100.7 & 1030$\pm$24\\
GC14 & 187.5766787 & 13.9705079 & 24.17$\pm$0.04 & 22.76$\pm$0.02 & 22.10$\pm$0.02 & 21.84$\pm$0.02 &  21.68$\pm$0.02 & 94.7 & 1035$\pm$33\\
GC15 & 187.6112049 & 13.9731678 & 24.28$\pm$0.04 & 22.91$\pm$0.02 & 22.33$\pm$0.02 & 22.05$\pm$0.02 &  21.95$\pm$0.02 & 41.6 & 1040$\pm$17\\
GC16 & 187.5983017 & 13.9758545 & 23.51$\pm$0.02 & 22.31$\pm$0.01 & 21.79$\pm$0.01 & 21.49$\pm$0.01 &  21.37$\pm$0.02 & 20.3 & 1088$\pm$13\\
N17$^{a}$ & 187.6018764 & 13.9803523 & 24.80$\pm$0.06 & 23.13$\pm$0.03 & 22.70$\pm$0.03 & 22.47$\pm$0.03 &  22.40$\pm$0.05 & 0.17 & 1066$\pm$20\\
GC21 & 187.6034917 & 13.9959557 & 23.86$\pm$0.03 & 22.57$\pm$0.01 & 22.01$\pm$0.02 & 21.75$\pm$0.01 &  21.57$\pm$0.02 & 56.6 & 1136$\pm$13\\
\hline
Star1 & 187.6339984 & 13.9077248 & 21.71$\pm$0.01 & 20.67$\pm$0.01 & 20.27$\pm$0.01 & 20.16$\pm$0.01 &  20.11$\pm$0.01 & 284.5 & 197$\pm$13 \\
Star2 & 187.6317882 & 13.9614617 & 24.45$\pm$0.04 & 22.42$\pm$0.01 & 21.40$\pm$0.01 & 21.19$\pm$0.01 &  20.98$\pm$0.01 & 124.7 & 20$\pm$15 \\
Star3 & 187.6353151 & 13.9254032 & 25.49$\pm$0.08 & 23.01$\pm$0.02 & 21.80$\pm$20.01 & 20.84$\pm$0.01 &  20.33$\pm$0.01 & 229.7 & -51$\pm$17 \\
\hline
\tablenotetext{a}{Nucleus of VC1287}
\enddata
\label{tab:data}
\end{deluxetable*}

We identify 7 objects with heliocentric velocities consistent with the Virgo cluster ($cz$ = 1079~\kms).
For these objects, we measure a systemic velocity $v_{\rm sys}$ = 1071$^{+14}_{-15}$ km/s,
from a maximum likelihood estimator (Hargreaves\etal1994).
6/7 objects are GCs associated with VCC~1287 (Table~\ref{tab:data}).
We identify ``N17'' as the nucleus of the galaxy, being both centrally located and
having a velocity ($1066 \pm 20$ km/s) close to the systemic velocity.
We calculate the maximum-likelihood line-of-sight velocity dispersion $\sigma_{\rm los}$ of the GCs
by excluding the nucleus,  $\sigma_{\rm los} = 33^{+16}_{-10}$ km/s within
8.1 kpc (the galactocentric radius of the outermost GC). Alternatively, if we include the nucleus
we obtain $v_{\rm sys}$ = 1071$^{+14}_{-15}$ km/s,  $\sigma_{\rm los} = 31^{+13}_{-9}$ km/s.
Here, we assume that our
measurement errors and the intrinsic system velocities have gaussian distributions.
The above uncertainties in the systemic velocity and velocity dispersion are
equivalent to those from the marginalized probability distributions.

By comparison, the  E + S0 galaxies in
the Virgo cluster have $\sigma_{\rm los}\sim$ 590 \kms, while dwarfs
(dEs + dS0s) have $\sigma_{\rm los}\sim$ 649 \kms (Binggeli\etal1985).
Therefore, these GCs are not associated with the cluster potential, but rather belong
to VCC~1287.
%We checked for rotation in the cluster system by performing non-linear
%fits to the cluster velocities as a function of position angle and found
%no statistically significant rotation in our spectroscopic sample.
%With the present data we cannot rule
%out dynamically significant rotation in this system \cite{Beasley2009}.

\section{Dynamical masses}

We use two approaches to estimate the gravitating mass
of VCC~1287, under the assumption that the GCs are in dynamical equilibrium.
One is to assume that the velocity dispersion of the GCs is
representative of the stars, allowing us to measure
the mass of the system at the half-light radius of the
galaxy. Specifically, we determine the ``half-mass'', $M_{1/2}$, at one galaxy-light
\re~(Wolf\etal2010):

\begin{equation}
M_{\rm 1/2} \simeq 930 \left(\frac{\langle\sigma_{\rm los}^2\rangle}{\rm{km^2~s^{-2}}}\right) \left(\frac{R_e}{\rm{pc}}\right)~{\rm M_{\odot}} . \label{eq:wolf}
\end{equation}

%where $\sigma_{\rm los}$ is the velocity dispersion, in this case that of the GC system. 
This gives a mass of 2.6$^{+3.4}_{-1.3}\times10^{9}$ \msun\
within 2.4~kpc.
%at one \re ($\sim2.4$ kpc from the galaxy center).
With $M_{g}=-13.3$, we calculate $L_g = 2.3\times10^{7}$~\lsun,
yielding a mass-to-light ratio in the $g$-band, $(M/L)_g = 106^{+126}_{-54}$ within
1~\re.

The second approach is to determine the mass within the radius of the outermost GC.
We use the ``tracer mass estimator'' (TME; Watkins\etal2010), which requires as input
the velocities of the tracers ($v_{\rm los}$), the projected galactocentric radius ($r$)
of the GCs to the outermost datum ($r_{\rm out}$), the slope of the gravitational potential
($\alpha$), the GC orbital (an)isotropy  ($\beta$)
and the power-law slope of the GC density profile ($\gamma$):

\begin{equation}\label{eq:watkins}
M = \frac{C}{G} \left<v_{\rm los}^2 r^{\alpha} \right>
\end{equation}
where
\begin{equation}
C = \frac{\left( \alpha + \gamma + 1 - 2 \beta\right)}
{I_{\alpha , \beta}} r_{\rm out}^{1- \alpha}
\label{eq:prefactor}
\end{equation}
and
\begin{equation}
I_{\alpha,\beta}
=\frac{\pi^{1/2}\Gamma(\frac{\alpha}{2}+1)}{4\Gamma(\frac{\alpha}{2}+\frac{5}{2})}
\left[\alpha+3-\beta(\alpha+2)\right]
%\left[\alpha(1-\beta)+3-2\beta\right]
\end{equation}
with $\Gamma(x)$ being the gamma function.  

Here we assume an isothermal potential ($\alpha=0$).
DM-only numerical simulations suggest an inner mass distribution
with $\alpha\simeq0.5$, whereas simulations that include baryons suggest $\alpha\simeq1.0$ (Schaller\etal2015).
Varying $\alpha$ based on these results changes the dynamical mass
by a few percent. We assume isotropic orbits ($\beta=0$). The assumption
of isotropy in the presence of radial ($\beta=0.5$) and tangential
($\beta=-3.0$) orbits causes mass over- and under-estimates, respectively, of $\sim10$\%.
These uncertainties are smaller than the statistical uncertainties arising from
the number of GC velocities measured.

We construct a surface density profile for the GC candidates identified from imaging
(with 6 confirmed spectroscopically, excluding the nucleus).
%This is shown in Fig.~\ref{fig:densGCs}.
Fitting a power-law ($N(r)\propto r^{-\gamma}$)  to this profile in the radial range
$0.5\leq r/$\re$ \leq 3.6$ gives $\gamma=1.2 \pm 0.4$\footnote{We measured the surface density profile, which goes
  into the TME prefactor as a volume density profile, i.e. $\gamma+1$ in equation~\ref{eq:prefactor}.}.
We also created a higher-SNR  stacked density profile from
19 dwarf elliptical (dE) galaxies in the ACS Virgo Cluster Survey (C\^ot\'e\etal2004) selected to 
have independent distance measurements (Mei\etal2007).
Each profile was normalised by the respective galaxy's \re  before stacking.
%We selected all GC candidates in these systems with a $p$GC~$\geq$~0.7 \cite{Jordan2004}.

A power-law fit to this profile gives $\gamma=1.5 \pm 0.1$.
Adopting this value in equation~\ref{eq:watkins} yields $M_{\rm TME}= (4.5 \pm 2.8) \times10^{9}$~\msun~ within 8.1 kpc.
The mass from the TME yields $(M/L)_g = 195 \pm 121$ within 8.1 kpc.
Note that the uncertainty in the slope of the surface density profile has a relatively small affect
on the inferred masses. For example, using $\gamma=1.2$ rather than $\gamma=1.5$
in equation~\ref{eq:watkins}~reduces the calculated mass by $\sim10$\%.

The above results indicate that both within 1~\re,
and within 8.1 kpc, VCC~1287 has substantial DM.
Using the relations of (Zibetti\etal2009), assuming a Kroupa initial mass
function, we calculate a stellar mass, $M_{\rm star} = (2.8 \pm 0.4) \times10^{7}$~\msun.
Therefore, within 1~\re~ the DM fraction in VCC~1287 is $\sim$~99\%.
Massive elliptical galaxies
and Virgo dEs typically have DM fractions of $<40$\%
within 1 \re (Cappellari\etal2006; Ry\'s\etal2014; Tortora\etal2016).
Only the faintest dwarf systems in the Local Group are known to have a similar
or higher DM fractions (e.g., Wolf\etal2010; McConnachie\etal2012)

\section{The virial mass}

\begin{figure*}[ht]
  \epsscale{1.1}
\plotone{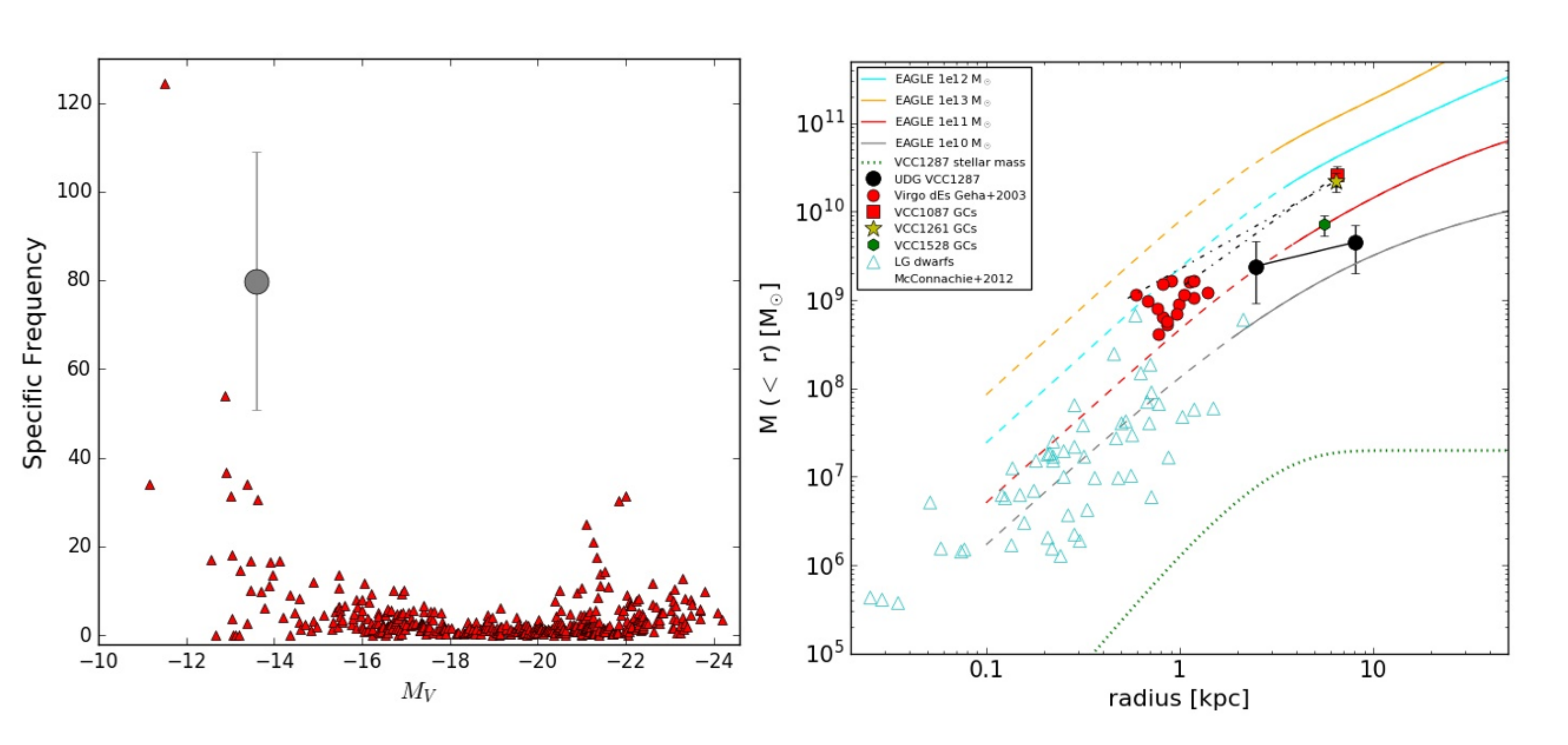}
\caption{Left panel: GC specific frequency of VCC~1287 compared to nearby galaxies (Harris\etal2013).
  VCC~1287 is a clear outlier with a very large number of GCs for its luminosity.
  Right panel: Mass of VCC~1287, Local Group dwarf galaxies (McConnachie\etal2012) and
  Virgo dEs (Geha\etal2003) compared to the cumulative mass profiles
  from EAGLE simulations (Schaller\etal2015).
  The measurements for VCC~1287 are not independent (see text). The long-dashed curves
  indicate the convergence radius for that halo mass.
  Short-dashed curves connect inner and outer mass measurements obtained for two
  Virgo dEs obtained from the stars and GC systems, respectively (Beasley\etal2009).
  The dotted curve shows the cumulative stellar mass profile for VCC~1287
  from our surface photometry.
}\label{fig:cumMASS}
\end{figure*}

Determination of the total (virial) mass of the DM halo 
gives insight into whether we are dealing with a dwarf or giant galaxy.
Since we cannot dynamically measure any mass beyond the radius
of the outermost GC (Newton's theorem),
we rely on numerical simulations to infer the halo mass of VCC~1287
using our observations as constraints.
We use the EAGLE simulations (Crain\etal2015; McAlpine\etal2015; Schaller\etal2015; Schaye\etal2015)
to determine the virial mass of the galaxy by comparing our mass measurements
to cumulative mass profiles of the simulations (Figure~\ref{fig:cumMASS}).
The EAGLE simulations include both DM and baryons, and therefore should give a realistic
representation of inner mass distributions of halos within the DM paradigm.
Using this approach we obtain $M_{200} = (8 \pm 4)\times10^{10}$ \msun,
from averaging the two mass measurements obtained from Figure~\ref{fig:cumMASS}.
Here, $M_{200}$ is the virial mass at the radius where
the density is 200 times the critical density of the universe.
This halo mass is similar to that reported for the
Large Magellanic Cloud (van der Marel \& Kallivayalil 2014).

To check this approach, we also ran spherically symmetric,
isotropic Jeans-mass models using {\it DM-only} mass distributions
and obtained $M_{200} = 5.5^{+24  }_{-4}\times10^{10}$~\msun,
in good agreement with our estimates from the EAGLE simulations.

Inferring virial masses is model-dependent, so
we sought an additional, independent check of our mass determinations.
Harris\etal(2013) obtained
the remarkable result that the ratio of the total mass in GCs in
a galaxy ($M_{\rm GCS}$) to the galaxy halo mass ($M_{\rm halo}$) is
constant over $\sim6$ decades in galaxy luminosity, specifically,
$M_{\rm GCS} / M_{\rm halo} = 6\times10^{-5}$. In other words, {\it the
  virial mass of a galaxy can be determined by
  totalling the mass in its GC system.}
We have already seen that VCC~1287 possesses a very large
number of GCs for its stellar luminosity (Figure~\ref{fig:cumMASS}).
%This in itself suggests that the galaxy may have a very massive halo
%for its stellar mass.
Using the relation of Harris\etal(2013),
with $22 \pm 8$ GCs, we obtain $M_{\rm halo}$ (GCs) = $(7.3 \pm 2.7)\times10^{10}$~\msun.
This halo mass is in excellent agreement with our dynamical inference.

\section{Discussion}

\begin{figure*}
\epsscale{1.2}
\plotone{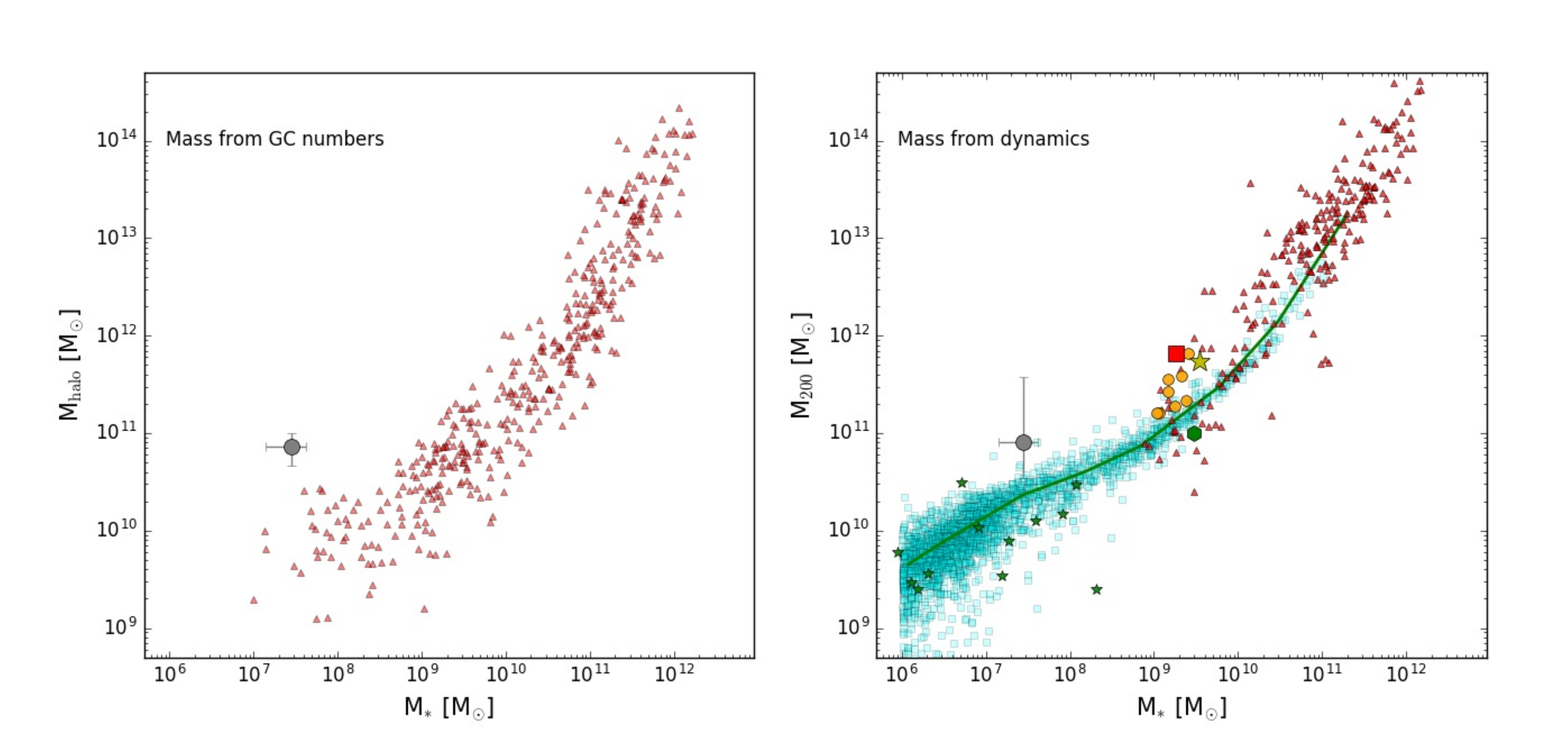}
\caption{
  Left panel: $M_{\rm star}$--$M_{\rm halo}$  relation for VCC~1287 (gray circle with error bars)
  compared to nearby galaxies (Harris\etal2013), with masses based on counting GCs.
  Stellar mass-to-light ratios were obtained from Zibetti\etal(2009),
  with $B-V$ colors from HyperLeda (green triangles).
  Right panel: $M_{\rm star}$--$M_{\rm 200}$ relation for VCC~1287 based on our dynamical masses
  compared to nearby galaxies (red triangles; Harris\etal2013), dEs (yellow circles; Geha\etal2003) 
  and Local Group dwarfs (green stars; McConnachie 2012). Masses for all these galaxies have been
  derived using the same methodology for which we derived the VCC~1287 mass.
  The dwarf-galaxy regime is sparsely sampled since we
  only show galaxies within at least 50\% of the EAGLE simulations convergence radius.
  Also shown are 2,582 central galaxies from EAGLE (cyan squares).
 }\label{fig:mstar_mhalo}
\end{figure*}

Much work in recent years has focused on the efficiency
of star formation in DM halos of a given
mass (e.g., Behroozi\etal2013; Moster\etal2013; Brook \& Di Cintio 2015; Tollet\etal2015).
One way to quantify this is using the
$M_{\rm star}-M_{\rm halo}$ relation. This relation is shown
in Figure~\ref{fig:mstar_mhalo} for our dynamically-based  $M_{\rm 200}$
and for the counting GC-based $M_{\rm halo}$ for VCC~1287 and selected galaxies.

Figure~\ref{fig:mstar_mhalo} shows that VCC~1287 is an outlier
independent of the method used to determine $M_{\rm halo}$. Its location
with respect to ``normal'' galaxies suggests that its stellar mass
is very low for its halo mass. In this context, VCC~1287 would need to have a stellar
mass a factor of $\sim100$ larger ($\sim5$ magnitudes brighter) in order to
be centred in these relations.
We measure a stellar fraction ($M_{\rm star} / M_{\rm halo}$) of $\sim3.5\times10^{-4}$,
whereas ``normal'' galaxies at this halo mass have $M_{\rm star} / M_{\rm halo} \sim 1\times10^{-2}$.

A halo-to-stellar mass ratio of $\sim$~3000 is unprecedented for any galaxy besides a dwarf spheroidal.
It suggests that galaxy formation is highly stochastic for halo masses of $\sim 10^{10}$--$10^{11} M_\odot$,
with stellar masses varying by factors of 100 or more at fixed halo mass.
This idea has been suggested as a solution to the problem of missing massive Milky Way satellites -- the
so-called ``too big to fail'' problem (Boylan-Kolchin\etal2011). 
However, the stochastic solution has been generally rejected as implausible, with extreme stochasticity
in simulations expected to set in at much lower masses (Brook \& Di Cintio 2015; Sawala\etal2015; Wheeler\etal2015).
The results for VCC~1287 raise the possibility that there are unidentified massive galaxies lurking at
still lower surface brightnesses, and that these DM halos are {\it not} too big to fail. 

Key questions are what is the formation history of this system, and is this
typical of all UDGs? We reject the idea that VCC~1287 is a tidal-dwarf system, as these
%These systems are believed to be the remnants of galaxy interactions and, as such, 
are expected to have total DM fractions of less than 10\% (Bournaud\etal2007).
%We obtain a DM fraction of $\sim99\%$.

We cannot presently rule out that this is a tidally stripped system, but this
interpretation is not favored by us.
We see no obvious tidal features in our imaging.
In addition, simulations suggest that in order to affect the stars and GCs,
more than 90\% of the DM must first be removed, and that the
more spatially extended GC system is affected before the stars (Smith\etal2015). 
We have shown that VCC~1287 has an unusually rich system of GCs for its stellar mass, so any tidal
mechanisms must preferentially remove stars over GCs. 
However, obtaining deeper imaging of this system in order to look for low surface brightness
tidal features would be very useful.

In a general sense this may be a ``quenched'' system (Boselli\etal2014):
perhaps a massive dwarf galaxy that had its star formation halted early due to
gas starvation as it fell into the Virgo cluster, and was therefore
unable to grow in stellar mass as it has in DM.
Assuming old ($\sim10$ Gyr) stars in VCC~1287, then the galaxy colors 
suggest low metallicities, [Fe/H]$\sim-1.5$ (Vazdekis\etal2010).
Such low metallicity is expected in mass-metallicity relations (e.g., Caldwell 2006),
and could be consistent with the early quenching of a dwarf-mass system.
However, we note that the galaxy might have a younger luminosity-weighted age, 
which we cannot rule out 
%and we are not able to constrain this satisfactorily 
with our present photometry.

To address the question of whether VCC~1287 is typical of UDGs, total
masses for more UDGs must be obtained. Determining stellar velocity dispersions for these faint systems
is expensive (or impossible) for 10m-class telescopes. Here we have demonstrated a more inexpensive route
through the kinematics of GC systems, at least for UDGs out to Virgo cluster distances.
%at distances less than or equal to that of the Virgo cluster.

However, we suggest the most efficient approach comes from simply counting
up the number of GCs in UDGs. We find excellent agreement between the halo masses inferred from dynamics and from
counting GCs for VCC~1287, and this approach offers an
inexpensive route to obtaining {\it virial masses for UDG systems out to
  Coma distances} (see also Mihos\etal2015).
For example, based on the calibrations of Harris\etal(2013), for ``dwarf-mass'' halos
($M_{\rm halo}\approx10^{10}$~\msun) we would expect  $\sim$5 GCs
in a UDG. For a quenched Milky-Way-like galaxy ($M_{\rm halo}\approx10^{12}$~\msun)
we would expect $\sim200$ GCs.

Relatively shallow $HST$ imaging (or deeper, multi-band ground-based imaging)
of these systems will allow for estimates of the total GC populations of UDGs, hence $M_{\rm halo}$,
and ultimately a better understanding of these galaxies.

\section{Acknowledgments}

We thank Giuseppina Battaglia for use of her maximum likelihood code, Claudio dalla Vecchia
for assistance with the EAGLE simulations and Jesus Falc\'on Barroso and Nacho Trujillo for useful
discussions. MAB and IMN were supported by the Programa Nacional de Astronom\'ia y Astrof\'isica
from the Spanish Ministry of Economy and Competitiveness, under grants  AYA2013-48226-C3-1-P,
AYA2013-48226-C3-2-P, AYA2013-48226-C3-3-P. MB acknowledges financial support from MINECO under
the 2011 Severo Ochoa Program MINECO SEV-2011-018.
AJR was supported by NSF grant AST-1515084.
DMD acknowledges support by the Sonderforschungsbereich (SFB) 881 of the German Research Foundation (DFG).
This article is based on observations made with the Gran Telescopio Canarias (GTC), in-
stalled in the Spanish Observatorio del Roque de los Muchachos of the Instituto de Astrof\'isica de Canarias,
on the island of La Palma. 
We acknowledge the Virgo Consortium for making their  simulation  data  available.  The EAGLE simulations
were  performed  using  the  DiRAC-2  facility  at Durham, managed by the ICC, and the PRACE facility Curie
based in France at TGCC, CEA, Bruy\`eres-le-Ch\^atel.

%% Facilities
% ===========================================================================
%{\it Facilities:} \facility{Nickel}, \facility{HST (STIS)}, \facility{CXO (ASIS)}.

%% Appendix
% ===========================================================================
%\appendix

%% References
% ===========================================================================

%% Figures
% ===========================================================================

%% End document
% ===========================================================================
\end{document}